\documentclass[11pt,oneside]{article}
\usepackage{setspace,graphicx,amssymb,amsmath,latexsym,amsfonts,amscd,amsthm,multirow,ctable,mathdots,caption,array,diagbox,mathtools}
\usepackage{tabularx,cite,mathrsfs}
\usepackage{authblk}
\usepackage{color}

\usepackage{fullpage}
\usepackage{stmaryrd}
\usepackage{rotating}

\usepackage{hyperref}

\newcommand{\bea}{\begin{eqnarray}}
\newcommand{\eea}{\end{eqnarray}}
\newcommand{\bee}{\begin{eqnarray*}}
\newcommand{\eee}{\end{eqnarray*}}
\newcommand{\al}{\begin{align*}}
\newcommand{\eal}{\end{align*}}
\newcommand{\be}{\begin{equation}}
\newcommand{\ee}{\end{equation}}

\newcommand{\bem}{\begin{pmatrix}}
\newcommand{\eem}{\end{pmatrix}}

\newcommand{\es}[2] {\begin{equation} \label{#1} \begin{split} #2 \end{split} \end{equation}}

\newcolumntype{R}{ >{$}r <{$}}
\newcolumntype{C}{ >{$}c <{$}}
\newcolumntype{L}{ >{$}l <{$}}
\newcolumntype{F}{>{\centering\arraybackslash}m{1.5cm}}

\newcommand{\mc}[1]{\mathcal{#1}}

\newcommand{\comment}[1]{}

\theoremstyle{definition}

\theoremstyle{remark}

\numberwithin{equation}{section}
\newtheorem*{eg}{Example}

\def\ba#1\ea{\begin{align}#1\end{align}}
\def\bg#1\eg{\begin{gather}#1\end{gather}}
\def\bm#1\em{\begin{multline}#1\end{multline}}
\def\bmd#1\emd{\begin{multlined}#1\end{multlined}}

\begin{document}

\setstretch{1.4}

\title{\vspace{-65pt}
\vspace{20pt}
    \textsc{
    An Extremal ${\cal N}=2$ Superconformal Field Theory
    }
}

\author[]{Nathan Benjamin
}
\author[]{Ethan Dyer}
\author[]{A. Liam Fitzpatrick}
\author[]{Shamit Kachru
}

\affil[]{Stanford Institute for Theoretical Physics, Department of Physics\\
and Theory Group, SLAC\\
Stanford University, Stanford, CA 94305, USA}

\date{}

\maketitle

\vspace{-1em}

\abstract{
We provide an example of an extremal chiral ${\cal N}=2$ superconformal field theory at $c=24$.  The
construction is based on a ${\mathbb Z}_2$ orbifold of the theory associated to the $A_{1}^{24}$
Niemeier lattice.  The statespace is governed by representations of the sporadic group $M_{23}$.

}

\clearpage

\tableofcontents


\section{Introduction}
\label{Introduction}

It is interesting to ask whether there are constraints on UV complete theories of quantum gravity.
In exploring this question, it is natural to start with the simplest such theory -- pure quantum
gravity with no matter fields.  While the question of existence of a theory with a given spectrum is beyond reach in most situations, AdS/CFT
provides a handle on this question in the context of asymptotically anti de Sitter quantum gravity.
For three dimensional AdS gravity, where the constraints of the extended conformal symmetry (Virasoro symmetry)
become especially strong, an exploration of this question was initiated in \cite{Ed}.

Perhaps the simplest tool to use in this context is modular invariance of the partition function of the dual 2d conformal field
theory.  Assuming the theory holomorphically factorizes, the case focused on in \cite{Ed}, 
the partition function is governed by a holomorphic function with $q$-expansion
\begin{equation}
\label{pf}
Z(\tau) = \sum_n c_n q^n,~~q = e^{2\pi i \tau}~.
\end{equation}
Factorized theories have central charge $c = 24 k$, and the expansion (\ref{pf}) describes a 
meromorphic function with a pole of order $k$ at $\tau = i\infty$.
Roughly speaking, the AdS/CFT map relates polar terms in the partition function to perturbative particle states, while
positive powers of $q$ have coefficients which capture the Bekenstein-Hawking entropy of AdS black holes.  

The definition of an extremal conformal theory -- a candidate dual to pure gravity -- advanced in \cite{Ed}, was that the polar
terms should be purely those arising from the stress-tensor multiplet together with its Virasoro descendants.  This allows one (with a few further
assumptions) to precisely fix a proposed partition function for quantum gravity at each value of $k$.
We note in passing that this is a strong assumption at large $c$; a much weaker criterion, fixing only coefficients up to
the power $q^{-a}$, with $a$ growing less quickly than $c$ as $c\to \infty$, would satisfy 
the conditions we can justifiably place on a theory of pure gravity as well.

In any case, at $k=1$, a conformal theory satisfying this constraint is available.  Remarkably, it is
the theory constructed by Frenkel, Lepowsky and Meurman in connection with Monstrous moonshine
some thirty years ago \cite{FLM}!  At higher values of $k$, which correspond to less negative values of
the cosmological term, precise candidates do not yet exist.  Various explorations resulting in
constraints of the possibilities
at these values appear in \cite{Yin,Matthias,Gaiotto,Matthiastwo}.\footnote{Related research on 3d gravity
partition functions \cite{Maloney}, general constraints on the gap in 2d CFTs \cite{Hellerman,Shapere}, and
the relationship of sparse CFT spectra with the emergence of spacetime \cite{HKS,us} has also appeared
in recent years.}

A similar discussion in \cite{Ed} of pure $\mathcal{N}=1$ supergravity in ${\rm AdS}_3$ yields a potential family of partition
functions at $c=12k$, and again, at $k=1$ a known superconformal theory fits the bill --
a ${\mathbb Z}_2$ orbifold of the supersymmetrized $E_8$ lattice
theory \cite{FLM, John}.

In \cite{GGKMO}, a criterion was similarly proposed for $\mathcal{N}=(2,2)$ superconformal theories dual
to pure extended supergravity in AdS$_3$.  These authors argued that the elliptic genus of the
conformal field theory, defined as
\begin{equation}
Z_{EG}(\tau,z) = {\rm Tr_{\mathcal{H}_{RR}}} (-1)^{F_L} e^{2 \pi i z J_0} q^{L_0-\frac{c}{24}} (-1)^{F_R} \bar q^{\bar L_0-\frac{c}{24}}~,
\end{equation}
is strongly constrained by the requirements of pure ${\cal N}=2$ supergravity in the bulk.  On general 
grounds, $Z_{EG}$ is a weak Jacobi form of weight $0$ and index $m$, where $m = c/6$.  These
authors were able to argue that for sufficiently large central charge, such extremal elliptic
genera do not exist; at small $m$, however, there are candidate Jacobi forms waiting to be matched
to actual $\mathcal{N}=2$ superconformal field theories.

Here, we propose that the case with $m=4$ -- of interest also because of the representation
theory visible in the $q$-expansion of the extremal elliptic genus, as we describe below -- can actually be realized.  In fact
the realization uses in a crucial way a known conformal field theory, that based on the
$A_1^{24}$ Niemeier lattice.  Extending the work of \cite{DGH}, we note that an enlarged, ${\mathbb Z}_2$ double cover of the $A_1^{24}/\mathbb{Z}_{2}$ orbifold admits an ${\cal N}=2$ superconformal symmetry, and in fact its chiral partition function matches the desired
extremal elliptic genus.  As our construction is chiral, this theory likely corresponds to a chiral gravity \cite{chiral} analogue
of the pure (super)gravity constructions envisioned in earlier works.  (For discussion of the supersymmetric extension of chiral gravity, see e.g. \cite{super}.)  As holomorphic factorization
of these gravity partition functions in any case engenders many confusions, it is plausible that 
the earlier two examples of extremal theories should also be viewed as duals of chiral gravity
theories in 3d.

The organization of this note is as follows. In \S\ref{sec:m4eg}, we describe the $m=4$ extremal elliptic genus.
In \S\ref{sec:a124}, we prove that an enlarged version of the $A_1^{24}$ Niemeier theory both admits
an ${\cal N}=2$ superconformal symmetry, and has a chiral partition function that reproduces the extremal elliptic genus. 
We close with some remarks in \S\ref{sec:remarks}.  Some important computations are relegated to an appendix.

\section{The extremal $\mc{N}=2$ partition function at $m=4$}
\label{sec:m4eg}

Elliptic genera, as well as chiral partition functions of ${\cal N}=2$ superconformal theories (with the extended algebra including a spectral flow symmetry \cite{Odake,Odake2}), are weak
Jacobi forms.  These are highly constrained objects; see the extensive discussion of their 
mathematical structure and their uses
in related problems in theoretical physics in \cite{DMZ}.

For our purposes, it suffices to know the following.  
A weak Jacobi form of index $m \in {\mathbb Z}$ and weight $w$ is a function $\phi(\tau,z)$ on 
${\mathbb H} \times {\mathbb C}$, satisfying
\begin{align}
\phi\left(\frac{a\tau+b}{c\tau+d}, \frac{z}{c\tau + d}\right) &= (c\tau + d)^w e^{2\pi i m \frac{cz^2}{c\tau + d}} \phi(\tau,z)~~~~~\begin{pmatrix} a&b \\ c&d \end{pmatrix} \in SL(2,\mathbb{Z})  \nonumber \\ 
\phi(\tau, z + \ell \tau + \ell^\prime) &= e^{-2\pi i m (\ell^2\tau + 2\ell z)}\phi(\tau,z)~~~~~~~~~~~~~~~~
\ell,\ell^\prime \in \mathbb{Z}~.
\end{align}

The ring of weak Jacobi forms of even weight is a polynomial algebra in four generators.
These are the Eisenstein series $E_4$
\begin{equation}
E_4(\tau) = 1 + 240 \sum_{n=1}^{\infty}\sigma_3(n)q^n = 1 + 240 q + 2160 q^2 + \cdots
\end{equation}
and $E_6$
\begin{equation}
E_6(\tau) = 1 -504 \sum_{n=1}^{\infty} \sigma_5(n) q^n = 1 - 504q - 16632 q^2 + \cdots
\end{equation}
of weights $4$ and $6$ and index $0$, as well as the Jacobi forms of $(w,m)$ equal to $(-2,1)$ and
$(0,1)$ given by
\begin{equation}
\varphi_{-2,1}(\tau,z) ={\theta_1(\tau,z)^2 \over \eta(\tau)^6} 
\end{equation}
and
\begin{equation}
\varphi_{0,1}(\tau,z) = 4 \left( {\theta_2(\tau,z)^2 \over \theta_2(\tau,0)^2} + {\theta_3(\tau,z)^2 \over
\theta_3(\tau,0)^2} + {\theta_4(\tau,z)^2 \over \theta_4(\tau,0)^2}\right)~.
\end{equation}

In terms of these functions, we can write the extremal elliptic genera of \cite{GGKMO} for low
values of $m$ as:
\begin{eqnarray}
Z_{EG}^{m=1} &=& \varphi_{0,1}\\
Z_{EG}^{m=2} &=& {1\over 6} \varphi_{0,1}^2 + {5\over 6} \varphi_{-2,1}^2 E_4\\
Z_{EG}^{m=3} &=& {1\over 48} \varphi_{0,1}^3 + {7\over 16} \varphi_{0,1}\varphi_{-2,1}^2 E_4
+ {13\over 24} \varphi_{-2,1}^3 E_6\\
Z_{EG}^{m=4} &=& {67\over 144} \varphi_{-2,1}^4 E_4^2 + {11\over 27} \varphi_{-2,1}^3 \varphi_{0,1} E_6 + {1\over 8} \varphi_{-2,1}^2 \varphi_{0,1}^2 E_4 + {1\over 432} \varphi_{0,1}^4~.\label{exteg}
\end{eqnarray}

We note in passing that the $m=2$ extremal elliptic genus arises as the chiral partition function of the
${\cal N}=2$ theory discussed in \S7 of \cite{six}.   That theory enjoys an $M_{23}$ symmetry.  Here, our focus will be on constructing an example of an extremal $m=4$ theory.\footnote{We note that it is plausible that more than one such theory exists, i.e. there could be additional constructions which also match the extremal $m=4$ elliptic genus.}

We now give a few more details about the would-be theory at $m=4$.

\begin{enumerate}
\item  The character expansion of the partition function is
\begin{eqnarray}
Z_{EG}^{m=4} &=& {\rm ch}^{N=2}_{{7\over 2};1,4} + 47 ~{\rm ch}^{N=2}_{{7\over 2};1,0} \nonumber \\
&+& (23 + 61984 q + \cdots) {\rm ch}^{N=2}_{{7\over 2};2,4}\nonumber\\
&+& (2024 + 485001 q + \cdots) ({\rm ch}^{N=2}_{{7\over 2};2,3} + {\rm ch}^{N=2}_{{7\over 2};2,-3}) \nonumber \\
&+& (14168 + 1659174 q + \cdots) ({\rm ch}^{N=2}_{{7\over 2};2,2} + {\rm ch}^{N=2}_{{7\over 2};2,-2}) \nonumber \\
&+& (32890 + 2969208 q + \cdots) ({\rm ch}^{N=2}_{{7\over 2};2,1}  + {\rm ch}^{N=2}_{{7\over 2};2,-1})~.
\label{pig}
\end{eqnarray}
Our conventions for characters are as in \cite{six}; ${\rm ch}^{N=2}_{l;h,Q}$ denotes the $\mathcal{N}=2$ superconformal
character with $l \equiv m-{1\over 2} = {c\over 6}-{1\over 2}$.  For BPS characters
$h={m \over 4} = {c\over 24}$, while for non-BPS characters  $h={m \over 4} + n$ with $n \in {\mathbb Z}$.  But because the non-BPS characters at various values of $n$ differ only by an
overall power of $q$, we write all of them as ${\rm ch}^{N=2}_{l;{m\over 4}+1,Q}$ and multiply
by the appropriate power of $q$ in front.

\medskip
\noindent
\item  The degeneracies in (\ref{pig}) are suggestive of an interesting (sporadic) discrete symmetry group.
The Mathieu group $M_{24}$ has representations of dimension 23 and 2024, for instance.  In fact
we will see below that this is no coincidence; our construction of an ${\cal N}=2$ SCFT with this chiral
partition function will be based on the $A_1^{24}$ Niemeier lattice, which enjoys $M_{24}$ symmetry.
The choice of ${\cal N}=2$ algebra breaks the symmetry group to $M_{23}$, which is the symmetry
group of the resulting ${\cal N}=2$ superconformal field theory.

\medskip
\noindent
\item  Because of the properties of the ring of weak Jacobi forms of weight $0$ and index $4$, such a 
form is determined by four coefficients in its $q,y$ expansion (where $y = e^{2\pi i z}$).  
For instance in the case at hand, the Ramond sector elliptic genus has a $q,y$ expansion
\begin{equation}
\label{hogpig}
Z_{EG}^{m=4,RR} = {1\over y^4} + 46 + y^4 + {\cal O}(q)~.
\end{equation}
Matching the coefficients of ${\cal O}(q^0)$ in (\ref{hogpig}) suffices to completely
determine this weak Jacobi form.
\end{enumerate}

\section{The $A_{1}^{24}/{\mathbb Z}_2$ orbifold}
\label{sec:a124}

The best known chiral conformal field theories are associated to theories of chiral bosons propagating
on even self-dual unimodular lattices of dimension $24k$, as well as their orbifolds.  At dimension 24,
there are precisely 24 such lattices -- the Leech lattice and the 23 Niemeier lattices \cite{CS}.  The chiral
conformal field theories at $c=24$ are conjecturally classified as well, starting with the work of 
Schellekens \cite{Schellekens}, as extended in \cite{Montague}.  For a recent review of progress in this
classification, see e.g. \cite{Lam}. 

We focus here on one of the theories associated to Niemeier lattices, the $A_1^{24}$ theory.
The $A_1^{24}$ Niemeier lattice contains the lattice vectors in the $A_1^{24}$ root lattice, as
well as additional vectors generated by the ``gluing vectors."  We discuss aspects of our construction in some detail below.  But first, we pause to give a general description of the theory, just by using simple facts
about the $A_{1}^{24}$ lattice.  These facts are as follows:

\begin{enumerate}
\item  The $SU(2)$ current algebra (associated to each of the $A_1$ factors) has three currents,
and so the $A_{1}^{24}$ theory is expected to have 72 states at conformal dimension $\Delta = 1$.
(This is correct in the full theory -- the additional gluing vectors do not add states at this low
conformal dimension.)

\medskip
\noindent
\item  This lattice theory, like all such theories, admits a canonical ${\mathbb Z}_2$ symmetry -- the one inverting the lattice vectors.  (In the language of 24 chiral bosons, it acts as $X^i \to -X^i$).   The
dimension of the ``twist fields" $\sigma^{a}$ creating the twisted sector ground states from the untwisted sector
is $\Delta_{\rm twist} = {3\over 2}$.  This is the right dimension for a supercharge; and following
the general construction of \cite{DGH}, in fact this theory can be promoted to an ${\cal N}=1$ superconformal
theory.  The total number of such twist fields is $2^{12}=4096$ \cite{Narain:1986qm}, and we will carefully choose two linear combinations of them to be the supercharges $G_\pm$.

\medskip
\noindent
\item  Furthermore, in the $A_{1}^{24}$ theory, one can actually promote to $\mathcal{N}=2$ superconformal invariance.
For an $\mathcal{N}=2$ superconformal algebra, we require an additional $U(1)$ current and a pair of supercharges
with charges $\pm$ under the $U(1)$.  For the $U(1)$ we select the $\mathbb{Z}_{2}$ invariant current,
\begin{equation}\label{firstcdef}
 J=2(e^{i\sqrt{2}X^{1}}+e^{-i\sqrt{2}X^{1}})\,,
\end{equation}
 in any of
 the 
24 copies of $SU(2)$ current algebra in the parent $A_1^{24}$ theory.  Its OPE with the dimension
$\frac32$ twist fields takes the form
\begin{equation}
J(z) \sigma^{a}(0) \sim {q^{ab}\over z} \sigma^b(0) + \cdots
\end{equation}
for some matrix $q^{ab}$, because $\sigma$ contains a twist operator for $X$ itself. This says that $\sigma$ is charged under the $U(1)$, and we will discuss how to diagonalize this action of $J$ on the $\sigma^a$'s. 

\medskip
\noindent
\item  With this choice of $U(1)$ generator, we can easily see how the lowest dimension states --
the 72 $\Delta = 1$ vectors -- appear in the superconformal theory.  Operators create NS or R
states depending on whether the singularities in their OPE with the supercharges are integral or
half odd integral.  
 It follows from the discussion in \cite{DGH} that
48 of the $\Delta = 1$ states arise in R sector, and 24 in the NS sector.  The former 48 are
the Cartan generators and the odd combinations of $\Delta = 1$ exponentials, while the latter
are the even combinations of exponentials.

From the normalization of the $U(1)$ current -- fixed by the ${\cal N}=2$ superconformal algebra to be
\begin{equation}
J(z) J(0) \sim {{c/3} \over z^2} + \cdots~,
\end{equation}
we can see that the $\Delta=1$ states have the following $U(1)$ charges.  The $W^{\pm}$ bosons of
the $SU(2)$ factor whose Cartan generator we have chosen have charges $\pm 4$.  All other states
at $\Delta = 1$ are neutral.

\medskip
\noindent
\item  Based on these observations, we can infer that the elliptic genus (or really, the chiral partition function) in the R sector takes the form
\begin{equation}
Z_{EG}^{m=4, RR}={1\over y^4} + 46 + y^4 + {\cal O}(q)~.
\end{equation}
This is enough to prove that it agrees with the extremal $m=4$ elliptic genus.
\end{enumerate}

We have used here the fact that $Z_{EG}$ is a weak Jacobi form.  To prove this, one needs not just the ${\cal N}=2$ superconformal algebra, but spectral flow invariance. This is guaranteed by the extended ${\cal N}=2$ algebra of Odake \cite{Odake,Odake2} and originally discussed in \cite{Lerche:1989uy}. A nice summary of this structure is provided
in \cite{Distler}.  In the case at hand, it suffices to exhibit additional chiral generators $\epsilon_{\pm}$ of dimension $\Delta = 4$ and $U(1)$ charges $\pm 8$.  To find such operators, consider the exponentials $e^{\pm i2\sqrt{2}X^{1}}$. These operators have dimension four and charge $\pm8$ under the appropriately normalized conventional Cartan generator $2\sqrt{2}i\partial X^{1}$  of $SU(2)$. In fact, they correspond to the highest and lowest weight states of a spin 2 $SU(2)$ representation, 
\es{hilo}{
|h,j,m\rangle &= |4,2,\pm2\rangle\,.
}
We are looking for states of definite charge under the current $J$ defined in (\ref{firstcdef}), not $2\sqrt{2}i\partial X^{1}$, but as the two currents are related by an $SU(2)$ rotation. It is not difficult to write down the eigenstates of $J$. The primary states with appropriate $J$ charge are thus given by:
\es{specflow}{
\epsilon_{\pm}&= |4,2,2\rangle\pm2|4,2,1\rangle+\sqrt{6}|4,2,0\rangle\pm2|4,2,-1\rangle+|4,2,-2\rangle\,.
}
These states are dimension four primaries of charge 8 and are invariant under the $\mathbb{Z}_{2}$ action which exchanges states of opposite spin. The states, $\epsilon_{\pm}$, correspond to operators in the $NS$ sector, and their action on the spectrum ensures spectral flow invariance.

We conclude that the double cover of the ${\mathbb Z_2}$ orbifold of the $A_{1}^{24}$ Niemeier lattice theory of
chiral bosons, has a chiral partition function given by the $m=4$ extremal elliptic genus.
More physically, it seems likely that the appropriate conjecture is that this theory describes
chiral ${\cal N}=2$ supergravity \cite{chiral,super} at an appropriate deep negative value of the cosmological constant.

We now provide a more detailed description of various elements of this theory.

\subsection{Review of the $A_{1}^{24}$ theory}
In this subsection, we briefly review the $A_{1}^{24}$ lattice and describe the enlarged theory, induced from the $A_{1}^{24}/\mathbb{Z}_{2}$ orbifold which realizes an $\mathcal{N}=2$ algebra with the extremal elliptic genus.

The one dimensional lattice, $A_{1}$ is a copy of $\mathbb{Z}$. The $A_{1}^{24}$ Niemeier lattice contains the direct sum of 24 copies of $A_{1}$ as a sub-lattice as well as additional lattice points specified by glue vectors. Explicitly, we can take basis vectors,
\es{fvecs}{
f_{1}&=(\sqrt{2},0,0,\ldots,0), \ \ \ f_{2} \, = \, (0,\sqrt{2},0,\ldots,0),  \ \ \ f_{24} \, = \, (0,0,\ldots,0,\sqrt{2})\, ,
}
and glue vectors,
\es{gvecs}{
g_{x_{1}x_{2}\ldots x_{24}}=\frac{1}{2\sqrt{2}}((-1)^{x_{1}},(-1)^{x_{2}},\ldots,(-1)^{x_{24}})\,,
}
where, $x_{i}$ take values $0$ and $1$ and the sequences, $\{x_{1},x_{2},\ldots,x_{24}\}$, run over words in the Golay code\footnote{Note this presentation makes the $M_{24}$ symmetry of $\Lambda$ manifest, as $M_{24}$ is the subgroup of $S_{24}$ which maps the Golay code to itself.}\cite{CS}. The full lattice consists of linear combinations,\footnote{In appendix \ref{app:twist} we give another presentation of the lattice, $\Lambda$.}
\es{latdef}{
\Lambda=\left\{\sum_{i}m_{i}f_{i}+\sum_{w}n_{w}g_{w} \ : \ m_{i},n_{w}\in\mathbb{Z} \ \& \ \sum_{w}n_{w}=0\right\}\,.
}

This lattice contains 48 length squared 2 vectors given by $\pm f_{i}$. The full lattice theta function is given by:
\es{LatticeTheta}{
\Theta_{\Lambda}(\tau)&=\sum_{\vec{v}\in\Lambda}q^{\vec{v}^{2}/2} \ = \ E_{4}(\tau)^3-\frac{21}{8} \theta_{2}(\tau)^8 \theta_{3}(\tau)^8\theta_{4}(\tau)^8\,.
}
The bosonic chiral CFT built from this lattice contains oscillator modes acting on primaries labeled by lattice vectors. In particular, in addition to the 48 dimension one operators coming from $\pm f_{i}$ the CFT contains an extra 24 of the form $\partial X^{i}$. The CFT partition function is given by:
\es{CFTPart}{
{\rm Tr}(q^{L_{0}-1})=\frac{\Theta_{\Lambda}(\tau)}{\eta(\tau)^{24}} \ = J(\tau)+72\,
}
where $J(\tau)$ is the modular invariant, normalized so that $J(\tau) = q^{-1} + \mathcal{O}(q)$ as ${\rm Im}(\tau) \to \infty$.

There is a natural $\mathbb{Z}_{2}$ action on bosonic CFTs given by $X^{i}\leftrightarrow-X^{i}$. One can obtain a new CFT as an orbifold by this $\mathbb{Z}_{2}$. For the case of CFTs built on Niemeier lattices, this orbifold operation provides an interesting map between theories. For the case of the Leech lattice, the analogous orbifolded theory is the Frenkel-Lepowsky-Meurman Monster module \cite{FLM,DGH}. It will be useful to consider a similar $\mathbb{Z}_{2}$ orbifold of the $A_{1}^{24}$.\footnote{The $A_{1}^{24}/\mathbb{Z}_{2}$ theory is equivalent to the theory on the Leech lattice, $\Lambda_{24}$. It is part of the sequence of theories, $(D_{16}E_{8}\ \textrm{or}\ E_{8}^{3})\rightarrow D_{8}^{3}\rightarrow D_{4}^{6}\rightarrow A_{1}^{24}\rightarrow \Lambda_{24}\rightarrow\mathbb{M}$, where each arrow represents a $\mathbb{Z}_{2}$ orbifold, mapping a theory with Coxeter number $h$ to $h/2-1$.}

Under the $\mathbb{Z}_{2}$ action $g$, the untwisted Hilbert space decomposes into $\mathbb{Z}_{2}$ invariant states and anti-invariant states:
\begin{eqnarray}
\mathcal{H}_{A_{1}^{24}}=\mathcal{H}_{+}^{+}+\mathcal{H}_{+}^{-}\,, 
\qquad  \left( g \psi = \pm \psi \quad \forall \psi \in \mathcal{H}_+^\pm \right).
\label{uthib}
\end{eqnarray}
There is also a twisted sector Hilbert space built on top of the dimension $3/2$ twisted sector vacuum.
\es{thib}{
\mathcal{H}_{-}=\mathcal{H}_{-}^{+}+\mathcal{H}_{-}^{-}\,.
}
The orbifolded theory corresponds to projecting onto only $\mathbb{Z}_{2}$ invariant states,
\es{orbhib}{
\mathcal{H}_{A_{1}^{24}/\mathbb{Z}_{2}}&=\mathcal{H}_{+}^{+}+\mathcal{H}_{-}^{+}\,.
}
The theory we wish to consider is neither the original $A_{1}^{24}$ theory, nor the orbifold, but rather the theory consisting of the enlarged Hilbert space of both invariant and anti-invariant Hilbert spaces in both twisted and untwisted sectors (as is standard in such constructions, see \cite{DGH}):
\es{orbhib2}{
\mathcal{H}_{\rm Double Cover}&=\mathcal{H}_{+}^{+}+\mathcal{H}_{+}^{-}+\mathcal{H}_{-}^{+}+\mathcal{H}_{-}^{-}\,.
}
In physics language, this full ${\cal H}_{\rm Double Cover}$ contains both the Neveu-Schwarz and Ramond sectors of the superconformal theory.
It is this theory\footnote{In a sense, we have taken the original theory $\mathcal{H}_{A_1^{24}}$, ``cut it in half" with the $\mathbb{Z}_2$ orbifold and then ``doubled it", then ``doubled it" again.} which possesses $\mathcal{N}=2$ supersymmetry and an extremal elliptic genus.
\subsection{$\mathcal{N}=2$ algebra, and matching the elliptic genus}
The chiral $\mathcal{N}=2$ algebra consists of the operators, $\{T,J,G^{\pm}\}$ satisfying the OPEs,
\es{n=2}{
T(z)T(w)&= \frac{c/2}{(z-w)^{4}}+\frac{2T(w)}{(z-w)^{2}}+\frac{\partial T(w)}{z-w}+\ldots\\
T(z)J(w)&= \frac{J(w)}{(z-w)^{2}}+\frac{\partial J(w)}{z-w}+\ldots\\
T(z)G^{\pm}(w)&= \frac{3}{2}\frac{G^{\pm}(w)}{(z-w)^{2}}+\frac{\partial G^{\pm}(w)}{z-w}+\ldots\\
J(z)J(w)&= \frac{c/3}{(z-w)^{2}}+\ldots\\
J(z)G^{\pm}(w)&=\pm\frac{G^{\pm}(w)}{z-w}+\ldots\\
G^{\pm}(z)G^{\mp}(w)&=\frac{2c/3}{(z-w)^{3}}+\frac{2J(w)}{(z-w)^{2}}+\frac{2T(w)+\partial J(w)}{z-w}+\ldots\\
G^{\pm}(z)G^{\pm}(w)&=\ldots
}

We would like to identify operators in our $A_{1}^{24}$ theory satisfying this $\mathcal{N}=2$ algebra. From our construction there are limited options. The only dimension $\frac32$ operators at our disposal are the $2^{12}$ twisted sector ground states, and so $G^{\pm}$ must be identified with two of these ground states. The remaining operators can be split into two sectors according to whether they have single-valued or double-valued operator product expansion with the supercurrent. These are the Neveu-Schwarz and Ramond sector respectively. From the perspective of our orbifold construction, the NS sector states consist of the $\mathbb{Z}_{2}$ invariant states in the untwisted sector and the $\mathbb{Z}_{2}$ anti-invariant states in the twisted sector.  The Ramond sector, in contrast, corresponds to anti-invariant states in the untwisted sector and invariant states in the twisted sector. 
\es{sectors}{
\text{NS}: &\ \ \mathcal{H}_{+}^{+}+\mathcal{H}_{-}^{-}\\
\text{R}: &\ \ \mathcal{H}_{+}^{-}+\mathcal{H}_{-}^{+}\,.
}
We will proceed to match the elliptic genus in the R sector above.

Focusing on the currents in our theory, we have 24 sets of three currents:
\es{currents}{
J^{i,0}(z)&=i\partial X^{i}(z)\\
J^{i,+}(z)&=\frac{1}{\sqrt{2}}\left(e^{i\sqrt{2} X^{i}}+ e^{-i\sqrt{2} X^{i}}\right)\\
J^{i,-}(z)&=\frac{1}{\sqrt{2}i}\left(e^{i\sqrt{2} X^{i}}- e^{-i\sqrt{2} X^{i}}\right)\,.
}
Of these three, only $J^{i,+}(z)$ is $\mathbb{Z}_{2}$ invariant, while the other two are anti-invariant.\footnote{Note, the $\pm$ in the superscript of $J^{i,\pm}$ denotes the $\mathbb{Z}_{2}$ charge. These are not the $SU(2)$ raising and lowering operators.} Thus in total, we have 24 NS sector currents, and 48 R sector currents. In the ${\cal N}=2$ algebra, the $U(1)$ current, $J(z)$, has a single valued OPE with $G^{\pm}(z)$ and thus must be a linear combination of the 24 $J^{i,+}(z)$. If we make identification,
 \es{defcurrent}{
 J(z)=2\sqrt{2}J^{1,+}(z)\,,
}
then the 48 Ramond sector states $\{ J^{i,0}, J^{i,-} |  i = 1, \dots 24\}$ decompose as $46$ neutral states (corresponding to $\{J^{i,0},J^{i,-}  \ | i>1\}$) and one charge $4$ state and one charge $-4$ state (linear combinations of $J^{1,0}$ and $J^{1,-}$).  This exactly matches the 
$\mc{O}(q^0)$ terms in the target elliptic genus (\ref{hogpig}), which in turn suffice to determine the full function.

This identification of the current also helps in finding the correct linear combination of the 4096 ground states, $\sigma^{a}$, to give the supercurrent, $G^{+}$. We are looking for some linear combination,
\es{supercurcombblah}{
G^{\pm}(z)&=\sum_{a=1}^{4096}v^{\pm}_{a}\sigma^{a}(z)\,,
}
with coefficients $v^{+}_{a}=(v^{-}_{a})^{*}$ chosen such that:
\es{supercurcomb}{
G^{+}(z)G^{-}(w)&=\frac{2c/3}{(z-w)^{3}}+\frac{2J(w)}{(z-w)^{2}}+\frac{2T(w)+\partial J(w)}{z-w}+\ldots\,.
}
The operators which can show up in the singular part of the OPE of any two twisted sector ground states are $\mathbb{Z}_{2}$ invariant, untwisted states of dimension less than three. Therefore, the only obstacle is choosing a linear combination of $\sigma^{a}$, such that only the current $J(z)$ and stress-tensor, and not the other $J^{i,+}(z)$, $i>1$ or spin two operators appear in the OPE. To accomplish this, it is useful to think about the charge of the twisted sector ground states under the 24 NS sector currents.

As we explain in appendix \ref{app:twist}, the 4096 $\sigma^{a}$ decompose into 2048 of charge $+1$ and 2048 of charge $-1$ under $J(z)$. Of the 2048 charge +1 states, which we will denote as $\sigma^{+\alpha}$, half are charge $+1$ and half charge $-1$ under any of the remaining $J^{i,+}(z)$, $i>1$. Thus, the definition,
\es{Gpropdef}{
G^{+}(z)&=\frac{1}{8\sqrt{2}}\sum_{\alpha=1}^{2048}\sigma^{+\alpha}(z), \ \ \ \ \ \ |G^{+}\rangle \, =\, \lim_{z\rightarrow0}G^{+}(z)|0\rangle\,.
}
guarantees the inner products,
\es{innerprod}{
\frac{\langle G^{-}|J^{i}_{0}|G^{+}\rangle}{\langle G^{-}|G^{+}\rangle} &=\frac{1}{2048}\sum_{\alpha=1}^{2048}q^{i\alpha} \, = \,\delta^{i1}\,,
}
which in turn implies the OPE, (\ref{supercurcomb}). Here, $q^{i\alpha}$ indicates the charge of $\sigma^{+\alpha}$ under the $i^{\text{th}}$ $U(1)$, and $\langle G^{-}|=(|G^{+}\rangle)^{\dagger}$. In appendix \ref{app:twist} we demonstrate that this choice of $G^{+}$ also ensures the decoupling of the unwanted dimension two currents. 

Note that this choice of supercurrent picks out one of 24 dimensions as special and thus supersymmetry only commutes with an $M_{23}\subset M_{24}$.

In summary, defining:

\es{bothdefs}{
T(z)&=-\frac{1}{2}\partial X^{i}\partial X_{i}(z)\\
J(z)&=2\left(e^{i\sqrt{2} X^{1}}(z)+ e^{-i\sqrt{2} X^{1}}(z)\right)\\
G^{+}(z)&=\frac{1}{8\sqrt{2}}\sum_{\alpha=1}^{2048}\sigma^{+\alpha}(z)\,.
}
we have an $\mathcal{N}=2$ algebra.  The chiral partition function of this ${\cal N}=2$ theory then matches the extremal elliptic genus (\ref{exteg}), and the theory has manifest $M_{23}$ symmetry.

\section{Remarks}
\label{sec:remarks}

There are several obvious questions that this construction suggests.

\noindent
$\bullet$ Can one find analogous constructions of extremal ${\cal N}=2$ theories at $c=18$, as well as at larger
values of $c$?  There is a proof in \cite{GGKMO} that extremal elliptic genera do not exist at 
sufficiently large $c$.  However, loosening the strict requirement (that all polar terms are generated
by ${\cal N}=2$ superconformal descendants of the gravity multiplet) to one which still applies to
almost all polar terms leaves scope for constructions at large $c$.

\noindent
$\bullet$ We now know of examples of such theories at $c=12$ (constructed in \cite{six}, where the connection to the extremal genus was not remarked on) and $c=24$.  In both cases, these
theories enjoy $M_{23}$ symmetry.  Is this a feature that would generalize to other examples?

\noindent
$\bullet$ All of the examples of extremal theories constructed to date enjoy sporadic group symmetries: there is the
bosonic theory with Monster symmetry, the ${\cal N}=1$ theory with Conway symmetry, and
these ${\cal N}=2$ theories with $M_{23}$ symmetry.  These groups enjoy interesting connections
with error correcting codes.  Is there a basic role for error correcting codes in quantum gravity?

\noindent
$\bullet$ More generally, it is of interest to construct chiral theories which pass various tests for admitting (perhaps stringy)
gravitational duals.  Criteria on elliptic genera of $\mc{N}=(2,2)$ theories were proposed in \cite{us}, and the same techniques would
lead to similar constraints on partition functions of chiral ${\cal N}=2$ theories.  As we know of no microscopic D-brane constructions which
yield purely chiral superconformal theories with gravity duals via near-horizon limits, finding such constructions is also
an attractive open problem.

\medskip
{\centerline{\bf{Acknowledgements}}
\medskip
We thank M. Cheng, J. Duncan, S. Harrison, A. Maloney, G. Moore, N. Paquette, E. Perlmutter, A. Shapere, D. Whalen, and especially R. Volpato for many discussions of closely related subjects.   We thank J. Duncan, S. Harrison, and G. Moore for helpful comments on a draft.   S.K. is grateful to the Aspen Center for Physics for hospitality when first reading about these theories in summer 2014, and when completing this paper in summer
2015.  N.B. and S.K. also thank the Perimeter Institute for hospitality, and the participants of ``(Mock) Modularity, Moonshine, and String Theory" for useful discussions about related subjects.
This research was supported in part by
the NSF via grant PHY-0756174, and the DoE Office of
Basic Energy Sciences through contract DE-AC02-76SF00515.  N.B. was also supported by a Stanford Graduate Fellowship.

\appendix
\section{Twisted Sector Ground States}\label{app:twist}

To verify the existence of the $\mathcal{N}=2$ algebra, and more fundamentally, to understand the structure of the twisted sector, $\mathcal{H}_{-}$, it is useful to elucidate some of the properties of the 4096 twisted sector ground states, $\sigma^{a}$.

As explained in \cite{Lepowsky,GDM}, the ground states $|a\rangle\equiv\lim_{z\rightarrow0}\sigma^{a}(z)|0\rangle$ form an irreducible representation of the untwisted sector operator algebra. More explicitly the action of any vertex operator corresponding to a lattice vector, $\lambda\in\Lambda$ can be defined as:
\es{OpAlg}{
\lim_{z\rightarrow0}(4z)^{\lambda^{2}/2}V_{\lambda}(z)|a\rangle=\gamma_{\lambda}|a\rangle\,.
}
Here the action of the vertex operator on the twisted sector ground state is encoded in the zero mode (cocycle factor) $\gamma_{\lambda}$.\footnote{The factor of 4 is required for mutual locality between the untwisted and twisted sector vertex operators \cite{GDM}. We will see that it is crucial in verifying that the supercurrent has the correct charge.} In order for the vertex operators $V_{\lambda}$ to be mutually local, we must have:
\es{locality}{
\gamma_{\lambda}\gamma_{\rho}=(-1)^{\lambda\cdot\rho}\gamma_{\rho}\gamma_{\lambda}\,,
}
and so the ground states, $|a\rangle$ fill out an irreducible representation of this algebra. This algebra is infinite dimensional, however all vectors in $2\Lambda$ give commuting operators, and so the non trivial part of the operator algebra is given by $\lambda\in\Lambda/2\Lambda$.

As is always the case for $\mathbb{Z}_{2}$ orbifolds, such as ours, the operators $\gamma_{\lambda}$ form a Clifford algebra and the ground states, $|a\rangle$ form the spinor representation of this algebra. In the remainder of this section we present the details of the Clifford algebra for the $A_{1}^{24}$ theory and show that the desired properties of the $\mathcal{N}=2$ algebra follow.
\subsection{Explicit Representation}

The description given above of the $A_{1}^{24}$ lattice, $\Lambda$, makes manifest the $M_{24}$ symmetry, however the constraints on the glue vectors are a little difficult to work with. We can instead view the $A_{1}^{24}$ lattice as the unrestricted span of a slightly different basis. We take the first 12 basis vectors to be the $f_{1},\ldots,f_{12}$ defined above. For the remaining 12 vectors, we take $v_{w}=\hat{v}_{w}/\sqrt{2}$, where $\hat{v}_{w=1\ldots12}$ is a basis of the Golay code.

The Golay code consists of 4096 words each consisting of 24 bits, with words containing 0, 8, 12, 16, or 24 ``1"s. The minimum Hamming distance between any two distinct words is 8. A convenient basis is given by the rows of a matrix, A,
\es{Golay}{
A = \left(
\begin{array}{cccccccccccccccccccccccc}
 1 & 1 & 0 & 1 & 0 & 1 & 0 & 1 & 0 & 1 & 0 & 1 & 1 & 0 & 0 & 0 & 0 & 0 & 0 & 0 & 0 & 0 & 0 & 0 \\
 1 & 1 & 1 & 0 & 1 & 0 & 1 & 0 & 1 & 0 & 1 & 0 & 0 & 1 & 0 & 0 & 0 & 0 & 0 & 0 & 0 & 0 & 0 & 0 \\
 0 & 1 & 1 & 0 & 0 & 1 & 1 & 1 & 1 & 1 & 0 & 0 & 0 & 0 & 1 & 0 & 0 & 0 & 0 & 0 & 0 & 0 & 0 & 0 \\
 1 & 0 & 0 & 1 & 0 & 0 & 1 & 1 & 1 & 1 & 1 & 0 & 0 & 0 & 0 & 1 & 0 & 0 & 0 & 0 & 0 & 0 & 0 & 0 \\
 0 & 1 & 0 & 0 & 1 & 0 & 0 & 1 & 1 & 1 & 1 & 1 & 0 & 0 & 0 & 0 & 1 & 0 & 0 & 0 & 0 & 0 & 0 & 0 \\
 1 & 0 & 1 & 0 & 0 & 1 & 0 & 0 & 1 & 1 & 1 & 1 & 0 & 0 & 0 & 0 & 0 & 1 & 0 & 0 & 0 & 0 & 0 & 0 \\
 0 & 1 & 1 & 1 & 0 & 0 & 1 & 0 & 0 & 1 & 1 & 1 & 0 & 0 & 0 & 0 & 0 & 0 & 1 & 0 & 0 & 0 & 0 & 0 \\
 1 & 0 & 1 & 1 & 1 & 0 & 0 & 1 & 0 & 0 & 1 & 1 & 0 & 0 & 0 & 0 & 0 & 0 & 0 & 1 & 0 & 0 & 0 & 0 \\
 0 & 1 & 1 & 1 & 1 & 1 & 0 & 0 & 1 & 0 & 0 & 1 & 0 & 0 & 0 & 0 & 0 & 0 & 0 & 0 & 1 & 0 & 0 & 0 \\
 1 & 0 & 1 & 1 & 1 & 1 & 1 & 0 & 0 & 1 & 0 & 0 & 0 & 0 & 0 & 0 & 0 & 0 & 0 & 0 & 0 & 1 & 0 & 0 \\
 0 & 1 & 0 & 1 & 1 & 1 & 1 & 1 & 0 & 0 & 1 & 0 & 0 & 0 & 0 & 0 & 0 & 0 & 0 & 0 & 0 & 0 & 1 & 0 \\
 1 & 0 & 0 & 0 & 1 & 1 & 1 & 1 & 1 & 0 & 0 & 1 & 0 & 0 & 0 & 0 & 0 & 0 & 0 & 0 & 0 & 0 & 0 & 1 \\
\end{array}
\right)\,,
}
with all words being the span of the basis vectors with coefficients in $\mathbb{F}_{2}$. In summary, the $A_{1}^{24}$ lattice can be expressed as:
\es{latdef2}{
\Lambda=\left\{\sum_{i=1}^{12}m_{i}f_{i}+\sum_{w=1}^{12}n_{w}v_{w} \ : \ m_{i},n_{w}\in\mathbb{Z}\right\}\,.
}
With this it is easy to give a description of $\Lambda/2\Lambda$. This consists of the $2^{24}$ vectors given by taking linear combinations of the 24 basis vectors with coefficients 0 or 1.
\es{lmod2l}{
\Lambda/2\Lambda=\left\{\sum_{i=1}^{12}m_{i}f_{i}+\sum_{w=1}^{12}n_{w}v_{w} \ : \ m_{i},n_{w}\in\mathbb{F}_{2}\right\}\,.
}
We are interested in a representation of the $2^{24}$ operators, $\gamma_{\lambda}$ for $\lambda\in\Lambda/2\Lambda$.

Though not clear in the above basis, there does exist a basis for $\Lambda/2\Lambda$, $\{e_{\mu=1\ldots24}\}$ where the operator algebra takes the form:
\es{cliff}{
\{\gamma_{e_{\mu}},\gamma_{e_{\nu}}\}&=2g_{\mu\nu}\,.
}
We prefer the basis consisting of $\{f_{i},v_{w}\}$, as the $f_{i}$ are simply related to 12 out of the 24 $\mathbb{Z}_{2}$ invariant $U(1)$ currents in the $A_{1}^{24}$ theory. Expression (\ref{cliff}) does make manifest, however, that the algebra formed by $\gamma_{\lambda}$ is the 24 dimensional Clifford algebra. This algebra has a unique irreducible representation of dimension $2^{12}$. And it is this representation under which the ground states transform. We choose to label them by the charges of the $12$ commuting matrices, $\gamma_{f_{i}}$.

Explicitly, we can take:
\es{gammamat}{
\gamma_{f_{1}}&=\sigma_{3}\otimes\mathbf{1}\otimes\mathbf{1}\otimes\ldots\otimes\mathbf{1}\,,\\
\gamma_{f_{2}}&=\mathbf{1}\otimes\sigma_{3}\otimes\mathbf{1}\otimes\ldots\otimes\mathbf{1}\,,\\
&\vdots\\
\gamma_{f_{12}}&=\mathbf{1}\otimes\mathbf{1}\otimes\mathbf{1}\otimes\ldots\otimes\sigma_{3}\,,\\
\gamma_{v_{1}}&=\sigma _2 \otimes\mathbf{1}\otimes \sigma _2 \otimes\mathbf{1}\otimes \sigma _2 \otimes\sigma _3 \otimes \sigma_{1} \otimes \sigma _3 \otimes \sigma_{1} \otimes \sigma _3 \otimes \sigma_{2} \otimes \sigma _1 \,,\\
\gamma_{v_{2}}&=\mathbf{1}\otimes\sigma _2 \otimes\mathbf{1}\otimes \sigma _2 \otimes\mathbf{1}\otimes \sigma _1 \otimes\sigma _3 \otimes \sigma_{2} \otimes \sigma _3 \otimes \sigma_{1} \otimes \sigma _1 \otimes\sigma _2 \,,\\
\gamma_{v_{3}}&=\mathbf{1}\otimes\mathbf{1}\otimes \sigma _2 \otimes \sigma_{2} \otimes \sigma _2 \otimes \sigma_{2} \otimes \sigma _1 \otimes\mathbf{1}\otimes \mathbf{1}\otimes \sigma_{2} \otimes \sigma _1 \otimes\sigma _3 \,,\\
\gamma_{v_{4}}&=\mathbf{1}\otimes \sigma_{2} \otimes \sigma _2 \otimes \sigma_{2} \otimes \sigma _2 \otimes \sigma_{1} \otimes \sigma _3 \otimes \sigma_{3} \otimes \sigma _2 \otimes\mathbf{1}\otimes \sigma _3 \otimes\sigma _1 \,,\\
\gamma_{v_{5}}&=\sigma _2 \otimes \sigma_{2} \otimes \sigma _2 \otimes \sigma_{2} \otimes \sigma _2 \otimes\mathbf{1}\otimes \sigma _3 \otimes \sigma_{2} \otimes \sigma _3 \otimes\mathbf{1}\otimes \sigma _2 \otimes\sigma _3 \,,\\
\gamma_{v_{6}}&=\sigma _2 \otimes \sigma_{2} \otimes \sigma _2 \otimes\sigma _2 \otimes\mathbf{1}\otimes \mathbf{1}\otimes \sigma_{2} \otimes \sigma _3 \otimes\sigma _3 \otimes \sigma_{1} \otimes \sigma _3 \otimes\sigma _1 \,,\\
\gamma_{v_{7}}&=\sigma _2 \otimes \sigma_{2} \otimes \sigma _2 \otimes\mathbf{1}\otimes \mathbf{1}\otimes\sigma _2 \otimes\mathbf{1}\otimes \mathbf{1}\otimes\sigma _1 \otimes \sigma_{2} \otimes \sigma _1 \otimes\sigma _3 \,,\\
\gamma_{v_{8}}&=\sigma _2 \otimes\sigma _2 \otimes\mathbf{1}\otimes \mathbf{1}\otimes\sigma _2 \otimes\mathbf{1}\otimes \mathbf{1}\otimes \sigma_{2} \otimes \sigma _2 \otimes \sigma _2 \otimes\mathbf{1}\otimes \sigma _2 \,,\\
\gamma_{v_{9}}&=\sigma _2 \otimes\mathbf{1}\otimes \mathbf{1}\otimes\sigma _2 \otimes\mathbf{1}\otimes \mathbf{1}\otimes \sigma_{2} \otimes \sigma _2 \otimes \sigma_{2} \otimes \sigma _2 \otimes \sigma_{2} \otimes\mathbf{1} \,,\\
\gamma_{v_{10}}&=\mathbf{1}\otimes\mathbf{1}\otimes \sigma _2 \otimes\mathbf{1}\otimes \mathbf{1}\otimes \sigma_{2} \otimes \sigma _2 \otimes \sigma_{2} \otimes \sigma _2 \otimes\sigma _2 \otimes\mathbf{1}\otimes \sigma _2 \,,\\
\gamma_{v_{11}}&=\mathbf{1}\otimes\sigma _2 \otimes\mathbf{1}\otimes \mathbf{1}\otimes \sigma_{2} \otimes \sigma _2 \otimes \sigma_{2} \otimes \sigma _2 \otimes\sigma _2 \otimes\mathbf{1}\otimes \sigma _2\otimes\mathbf{1} \,,\\
\gamma_{v_{12}}&=\sigma _2 \otimes\mathbf{1}\otimes \mathbf{1}\otimes \sigma_{2} \otimes \sigma _2 \otimes \sigma_{2} \otimes \sigma _2 \otimes\sigma _2 \otimes\mathbf{1}\otimes \mathbf{1}\otimes\mathbf{1}\otimes \sigma _2 \,.
}
With this choice the ground states transform in the $2^{12}$ dimensional spinor labeled by:
\es{gs}{
|a\rangle&=|\pm\pm\ldots\pm\rangle\,.
}
\subsection{Supercurrent}
In \S\ref{sec:a124}, we defined the supercurrent as the state:
\es{supercur}{
|G^{+}\rangle&=\frac{1}{8\sqrt{2}}\sum_{\{i_{2\ldots12}\}}|+i_{2}i_{3}\ldots i_{12}\rangle\,.
}
We can now explicitly check some of the properties of $G^{+}$. Firstly, the $U(1)$ charge is given by:

\es{U(1)charge}{
\frac{\langle G^{-}|J| G^{+}\rangle}{\langle G^{-}|G^{+}\rangle}&=\frac{1}{2}\frac{\langle G^{-}|\gamma_{f_{1}}+\gamma_{-f_{1}}| G^{+}\rangle}{\langle G^{-}|G^{+}\rangle}\\
&=\frac{\langle G^{-}|\gamma_{f_{1}}| G^{+}\rangle}{\langle G^{-}|G^{+}\rangle}\\
&=1\,.
}
Indeed the ground state has the correct charge, note that the factor of 4 in (\ref{OpAlg}) is crucial in verifying that the ground state is charge 1.

For the other currents, it is easy to see that the three point interaction vanishes.
\es{vanishing}{
\langle G^{-}|J^{i\neq1}| G^{+}\rangle &=0\,.
}
This is due to the cancellation of terms with positive and negative charge in the sum, (\ref{supercur}).

It is also possible to see that the supercurrent does not couple to any spin-two currents other than the stress tensor. To check this, note that the dimension two operators take one of the following three possible forms:
\begin{enumerate}
\item $\partial X^{i} \partial X^{j}$,
\item $e^{i\lambda\cdot X}$, For $\lambda$ of the form $\lambda=f_{i}+f_{j}$,
\item $e^{i\lambda\cdot X}$, For $\lambda$ of the form $\lambda=\sum_{w}n_{w}v_{w}$, with $\lambda^{2}=4$.
\end{enumerate}

The coupling of operators of the first type to the supercurrent can be computed directly by expanding $\partial X^{i}$ in modes. 
\es{modeexp}{
\partial X^{i}(z)&=-i\sum_{r\in\mathbb{Z}+1/2}\frac{c^{i}_{r}}{z^{r+1}} \, , \ \ \ \ \ \ c^{i}_{r}|a\rangle\,=\, 0 \ \ \ \forall \,r\,>\,0\, , \ \ \ \ \ \ [c_{r}^{i},c_{s}^{j}]\,=\,r\delta^{ij}\delta_{r+s,0}
}
The matrix of operators, $\partial X^{i}\partial X^{j}$, can be separated into a symmetric traceless component, an antisymmetric component, and a trace. As the $c_{r}^{i}$ commutation relations are proportional to $\delta^{ij}$ only the trace gives a non vanishing contribution:
\es{partops}{
\langle G^{-}|\partial X^{i}\partial X^{j}| G^{+}\rangle = \frac{1}{24}\delta^{ij}\sum_{k}\langle G^{-}|\partial X^{k}\partial X^{k}| G^{+}\rangle\,,
}
but this is exactly the contribution of the stress tensor.

For the second type of operator, we can write the three point interaction with the supercurrent as:
\es{jjgg}{
\frac{1}{4}\langle G^{-}|e^{i\lambda\cdot X}| G^{+}\rangle&=\langle G^{-}|\gamma_{\lambda}| G^{+}\rangle\\
&=\langle G^{-}|\gamma_{f_{i}+f_{j}}| G^{+}\rangle\\
&=\pm\langle G^{-}|\gamma_{f_{i}}\gamma_{f_{j}}| G^{+}\rangle\\
&=0\,.
}
The vanishing in the last line is the same argument as for the 24 currents. Contributions with positive and negative charge cancel out of the sum. 

The vanishing of the coupling with the third class of operators is slightly more subtle. In the Golay code there are, 759 words of length eight (octets). Of these, 253 have a 1 in the first column, and thus lead to operators with anti-commute with $J$. Such operators must have a vanishing $G^{+}G^{-}\mathcal{O}$ coupling, as $G^{+}$ has definite charge under $J$. The remaining 506 octets can be further decomposed under the subgroup of $M_{24}$ which fixes the first column, $M_{23}$. The state $|G^{+}\rangle$ is a singlet under $M_{23}$ and so can only couple to singlets. Of the $506$ dimension-two operators, there is only one singlet \cite{Atlas}, and thus the other $505$ dimension-two operators automatically decouple.  We do not know of an elegant argument for the decoupling of the singlet. One can explicitly check however, using the basis of gamma matrices (\ref{gammamat}), that in fact all of the 759 dimension two operators of this form have vanishing three point interaction with $G^{+}G^{-}$.

\bibliographystyle{JHEP}
\bibliography{refs}
\end{document}